\documentclass[%
 reprint,
 amsmath,amssymb,
 aps,
prl,
floatfix,
]{revtex4-2}

\usepackage{graphicx}
\usepackage{dcolumn}
\usepackage{bm}%

\usepackage{framed}
\usepackage{xcolor}
\colorlet{shadecolor}{blue!20}

\begin{document}

	\preprint{APS/123-QED}
	\title{ Magnetic devil's staircase in UAgBi$_{2}$}
	
	\author{ G. S. Freitas$^{1,2}$,  K. Kirchner$^3$,  C. Girod$^{1}$, D. R. Yahne$^{3,4}$, W. Simeth$^{1}$,  C. S. T. Kengle$^{1}$, F. B. Carneiro$^{1,5}$, P. G. Pagliuso$^{2}$, J.D. Thompson$^{1}$, M. Janoschek$^{4,6}$, O. Zaharko$^4$, S.M. Thomas$^{1}$, and P. F. S. Rosa$^{1}$}

	\affiliation{
		$^{1}$ Los Alamos National Laboratory, Los Alamos New Mexico, USA.\\
		$^{2}$ Instituto de F\'isica Gleb Wataghin, UNICAMP, Campinas-SP, 13083-859, Brazil.\\
        $^{3}$ PSI Center for Neutron and Muon Sciences CNM, 5232 Villigen PSI, Switzerland\\
		$^{4}$ Neutron Scattering Division, Oak Ridge National Laboratory, Oak Ridge, Tennessee 37831, USA\\
        $^{5}$ Centro Brasileiro de Pesquisas F\'isica, 22290-180, Rio de Janeiro, RJ, Brazil\\
        $^{6}$ Physik-Institut, Universität Zürich, Winterthurerstrasse 190, CH-8057 Zürich, Switzerland
        }

	\begin{abstract}
		
		Materials characterized by competing interactions often exhibit a large number of nearly degenerate periodic states. Here we show that layered UAgBi$_2$ hosts a cascade of field- and temperature-induced magnetic transitions. Based on specific heat, thermal expansion, and neutron diffraction, we construct a phase diagram that reveals at least seven nearly degenerate magnetic states in UAgBi$_2$. The observed multi-step magnetization process can be understood by square-wave structures with distinct propagation vectors $\mathbf{k}$=(0, 0, \textit{k}) in the presence of strong easy-axis anisotropy that confines the moments along the \textit{c} axis. Our findings are consistent with a magnetic devil's staircase described by the axial next-nearest neighbor Ising (ANNNI) model and place UAgBi$_2$ as a rare realization of the devil's staircase in a 5\textit{f}-electron system.
	
	\end{abstract}

	\maketitle

	Periodic modulated patterns are ubiquitous in physical, chemical, and biological systems that host competing degrees of freedom \cite{seul_domain_1995}. Examples range from Cooper pair modulations at the atomic scale in superconductors \cite{hamidian_detection_2016,kong_cooper-pair_2025} to steady-state mesoscale Turing patterns in chemical reactions \cite{castets_experimental_1990,ouyang_transition_1991}. In magnetic materials, competing exchange interactions can frustrate conventional order and generate multiple nearly degenerate magnetic configurations \cite{schmidt_frustrated_2017,lacroix_frustrated_2010,diep_frustrated_2013,lacroix_introduction_2011}. Although magnetic frustration is often associated with the suppression of conventional order, as in quantum spin liquids \cite{broholm_quantum_2020}, it can also stabilize highly tunable ordered phases with distinct magnetic modulations. A particularly striking example is the devil's staircase, in which a sequence of distinct magnetic configurations is stabilized as an external parameter, such as magnetic field or temperature, is varied \cite{bak_ising_1980,bak_commensurate_1982,bak_devils_1986}.

	Field-induced devil's staircase behavior has been observed in many classes of materials with $4f$ and $3d$ elements, including CeSb \cite{kuroda_devils_2020}, CeSbSe \cite{chen_possible_2017}, CeRh$_3$Si$_2$ \cite{pikul_giant_2010},  TbNi$_2$(Ge, Si)$_2$ \cite{islam_neutron_1998,shigeoka_metamagnetism_1992}, SrCo$_6$O$_{11}$ \cite{matsuda_observation_2015}, and SrCu$_2$(BO$_3$)$_2$ \cite{takigawa_incomplete_2013}.
	CeSb is a particularly well-established case of magnetic devil’s staircase (MDS) behavior wherein the combination of single-ion anisotropy and competing exchange interactions leads to a series of modulated phases that are commensurate with the underlying lattice and vary with temperature and applied field \cite{kuroda_devils_2020}. 
	The MDS behavior in CeSb was theoretically understood, at least at a qualitative level, half a century ago by the so-called axial next-nearest neighbor Ising (ANNNI) model: a spin-1/2 Ising model in $d$ dimensions with ferromagnetic (FM) coupling within $(d-1)$-dimensional layers but competing FM and antiferromagnetic (AFM) interactions between nearest and next-nearest layers along one, unique spatial axis perpendicular to the layers. Even at the mean field level, the self-consistent solution of the ANNNI model uncovers multiple phase transitions between commensurate phases, and the number of phases can be increased by tuning the exchange parameters \cite{fobesTunable2018}. 
    
    Given the general form of the ANNNI Hamiltonian, MDS behavior was originally expected to be a common observation in local-moment systems; however, there are only a few cases of 5\textit{f}-based materials for which this possibility has been considered \cite{lander_magnetic_1995,burlet_neutron-diffraction_1984,langridge_resonant_1994,fobes_realization_2017,kawarazaki_frozen-devils_1994}. For example, a nearly continuous variation of the ordering wavevector (0, 0, $k$) is observed in U$_3$Al$_2$Ge$_3$ below its AFM transition before a first-order transition to a FM ground state takes place \cite{fobes_realization_2017} -- a behavior well described by the ANNNI model. In U(Ru$_{1-x}$Rh$_x$)$_2$Si$_2$, the presence of multi-\textit{k} phases as a function of temperature led to an initial proposal that the multiple states stem from the ANNNI model \cite{kawarazaki_frozen-devils_1994}, but this scenario remains unconfirmed.
	
	In this Letter, we report the discovery of UAgBi$_2$, a layered compound with a UBi$_2$-Ag-UBi$_2$ stacking structure that has the necessary ingredients to give rise to a rare realization of the devil's staircase in a $5f$-electron material. The incorporation of Ag layers in UBi$_{2}$ introduces significant changes to the U single-ion anisotropy and tunes the $c$-axis magnetic exchange interactions towards the ANNNI model. Our specific heat data also reveal a small Sommerfeld coefficient of $\gamma = 40$~mJ/mol.K$^{2}$, which indicates that $5f$ electrons in UAgBi$_2$ are close to the localized limit wherein MDS behavior is more likely to arise. Finally, our thermodynamic and neutron diffraction measurements reveal a complex magnetic phase diagram in UAgBi$_2$ with the expected fingerprint of multi-step MDS behavior, which can be understood by a sequence of distinct (0, 0, \textit{k}) square-wave structures.

	\begin{figure}[!ht]
		\includegraphics[width=0.95\columnwidth]{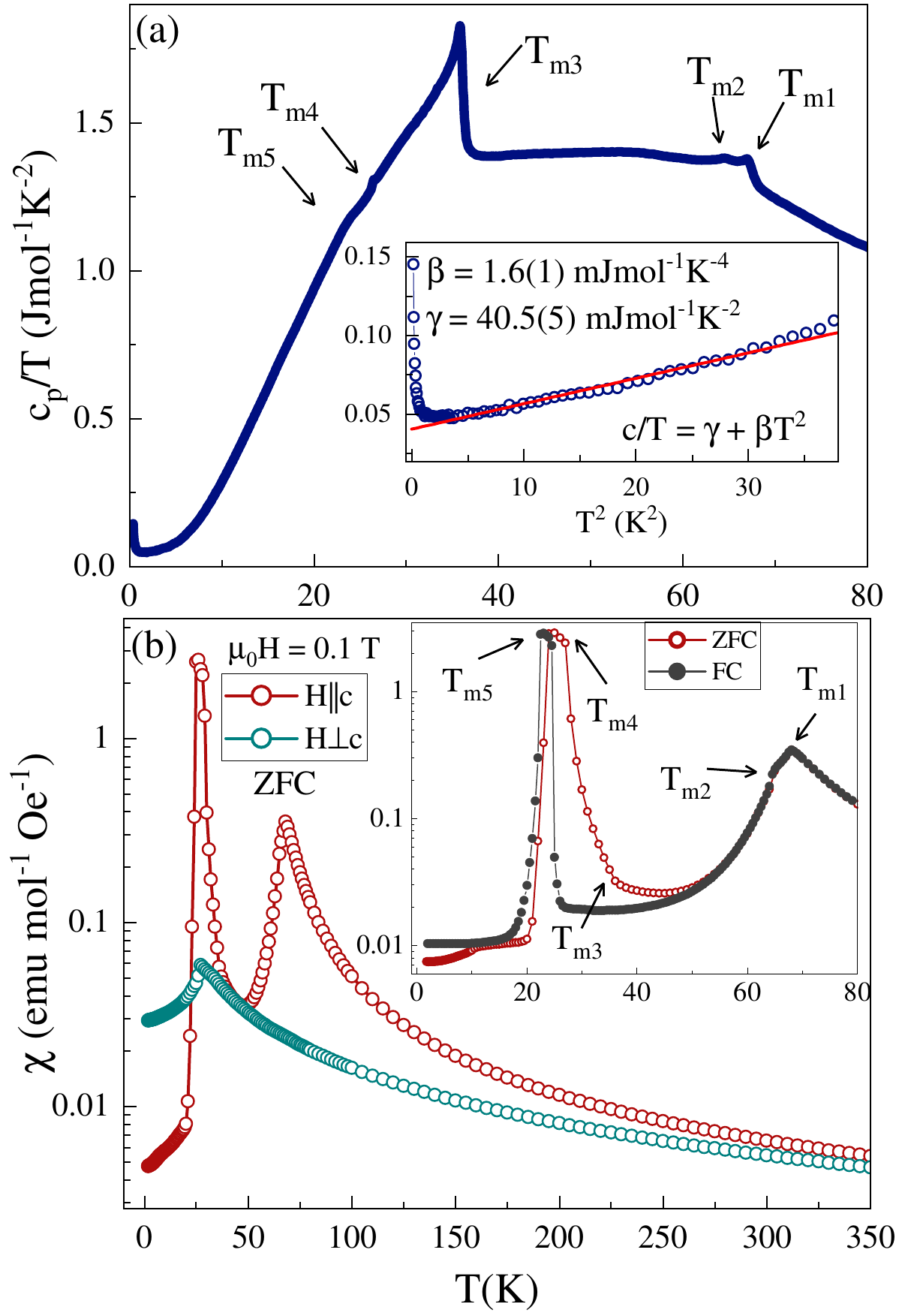}
		\caption{ (a) Zero-field specific heat $c_{p}(T)/T$ of UAgBi$_2$ as a function of temperature from 0.4 to 80 K. The inset shows a zoom of the low-temperature behavior plotted as a function of $T^2$; the red line represents the fit discussed in the text. (b) ZFC magnetic susceptibility of UAgBi$_2$ measured under an applied field of 0.1 T parallel (red open symbols) and perpendicular (teal open symbols) to the $c$-axis. The inset displays a zoomed-in view of the ZFC (red open symbols) and FC curves (black solid symbols) for H$||$c. }
		\label{cpxMxT}
	\end{figure}

Our single crystal x-ray diffraction measurements confirm that UAgBi$_{2}$ crystallizes in nonsymmorphic space group $P4/nmm$ (\#129) with similar atomic positions to other members of the 112 family. 
The $c$ axis in UAgBi$_{2}$, however, is 10~\% larger than that in UCuBi$_{2}$ and 7\% larger than that in UAuBi$_{2}$ \cite{kaczorowski_structural_1992,rosa_ferromagnetic_2015,long_Magnetic_2025}. This non-monotonic variation is important to the unique properties of UAgBi$_{2}$, as discussed later. The structural parameters and atomic positions of UAgBi$_{2}$ at room temperature are shown in Table~\ref{table2} in the Appendix. 

In contrast to single-crystalline UCu$_x$Bi$_2$, for which a significant Cu-site vacancy was reported \cite{long_Magnetic_2025}, our single-crystal x-ray diffraction refinement of UAgBi$_2$ indicates a fully occupied Ag site within the uncertainty of the refinement, as shown in Table~\ref{table2}. Following a strategy similar to that used for CeAgBi$_2$ \cite{thomas_hall_2016}, a starting ratio of 1:10:10 was used during crystal growth to reduce the likelihood of Ag vacancies.

	Figure~\ref{cpxMxT}(a) shows the temperature dependence of the specific heat, $c_{p}(T)$, measured on warming, wherein five transitions ($T_{m}$) are clearly observed at $T_{m1}= 67$~K, $T_{m2}= 64$~K, $T_{m3}= 36$~K, $T_{m4}= 27$~K, and T$_{m5}$ = 24~K. To extract the electronic Sommerfeld coefficient ($\gamma$), we perform a linear fit to $c_{p}(T)$ $vs$ $T^{2}$ at low temperatures. As shown in the inset of Figure \ref{cpxMxT}(a), we obtain $\gamma = 40$~mJ/mol.K$^{2}$, consistent with weak Kondo renormalization. For comparison, $\gamma = 75$~mJ/mol.K$^{2}$ in UAuBi$_{2}$ \cite{rosa_ferromagnetic_2015}, which suggests that the $5f$ electrons in UAgBi$_{2}$ are more localized.

	Figure~\ref{cpxMxT}(b) shows magnetic susceptibility ($\chi$) as a function of temperature with a magnetic field of 0.1~T applied parallel ($\chi_{||}$) and perpendicular ($\chi_{\perp}$) to the $c$ axis. Under zero-field cooling (ZFC) conditions, all five anomalies detected in $c_{p}(T)$ data are evident in $\chi_{||}$, wheres only $T_{m5}$ is clearly observed in $\chi_{\perp}$. The larger $c$ axis susceptibility and the lack of features for in-plane fields suggest that the uranium moments point primarily along the $c$ axis and that the $c$ axis is the magnetic easy axis, similar to most members of the 112 family \cite{kaczorowski_structural_1992,rosa_ferromagnetic_2015,adriano_physical_2014,piva_electronic_2020,thomas_hall_2016}.

	\begin{figure*}[!ht]
		\includegraphics[width=1\textwidth]{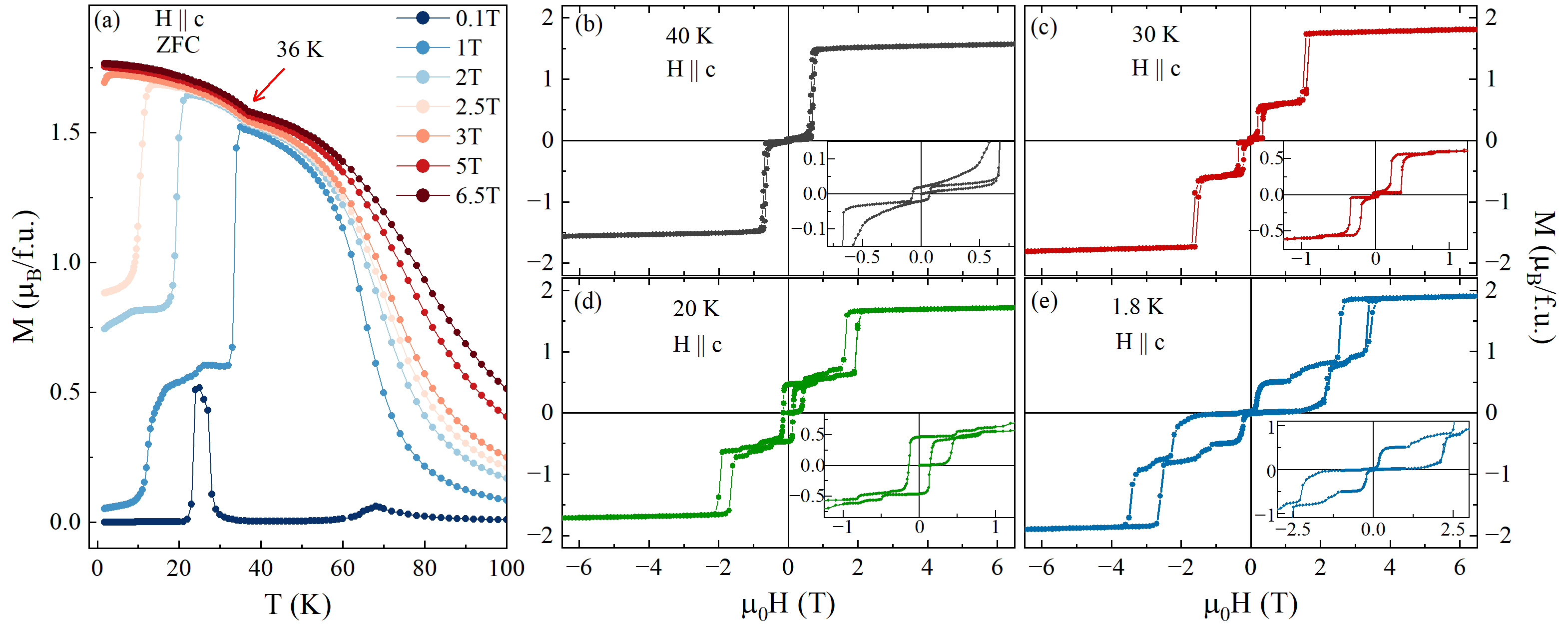}
		\caption{ (a) Zero-feld-cooled magnetization as a function of temperature under different applied magnetic fields along the c axis. (b)-(e) Magnetization as a function of applied magnetic fields at different temperatures for fields parallel to the c axis. }
		\label{MxTxH}
	\end{figure*}
	
	Curie-Weiss fits to $\chi(T)$ data from 200 K to 350 K yield an effective moment of $\mu_{eff} = 3.5(1)~\mu_{B}$ [3.6(1)~$\mu_{B}$] for fields applied parallel [perpendicular] to the $c$ axis, which match well the expected Hund's moments for both U$^{3+}$ (3.62 $\mu_{B}$) and U$^{4+}$ (3.58 $\mu_{B}$). Thus, the uranium valence cannot be determined based on these fits alone. A Curie-Weiss temperature of $\theta_{CW} = 68.0(5)$~K is extracted for fields along the $c$ axis, whereas $\theta_{CW} = -3.4(3)$~K for in-plane fields. In the simplest approximation using a molecular field in the absence of crystalline electric field (CEF) effects, the anisotropy of $\theta_{CW}$ points to two effective exchange interactions with opposite signs, $i.e.$, FM interactions along the $c$ axis and AFM interactions in the $ab$ plane \cite{adriano_physical_2014,piva_electronic_2020,leciejewicz_neutron-diffraction_1967}. This conclusion, however, is at odds with the anisotropy of interactions commonly observed in the family $RMX_{2}$ ($R =$ lanthanide or actinide, $M=$ transition metal, and $X=$ Sb, Bi), $i.e.$, dominant FM interactions in the $ab$ plane. X-ray and neutron magnetic diffraction measurements have consistently shown FM planes that are antiferromagnetically aligned. Our results thus suggest that CEF effects are likely relevant in UAgBi$_2$.

	We now turn to field-dependent magnetization data. Figures~\ref{MxTxH}(a) show the magnetization as a function of temperature [$M(T)$] at different fields applied parallel to the $c$ axis.  When a field of only 1~T is applied along the $c$ axis, the AFM-like transitions, T$_{m1}$ and T$_{m2}$, combine into a single, broader FM-like feature that moves to higher temperatures in field. Also at 1~T, two distinct transitions emerge around 34~K and 36~K in $M(T)$. As field is further increased, the transition at 34~K is suppressed, in agreement with AFM behavior, whereas the transition at 36~K (T$_{m3}$) is clearly observed under all applied fields and reveals an upturn in $M(T)$. These results indicate the presence of a FM contribution to T$_{m3}$.

	Figures~\ref{MxTxH}(b)-(e) show magnetization loops at different temperatures. At 40~K, only one field-induced transition is observed when field is applied along the easy $c$ axis [Figure~\ref{MxTxH}(b)].    
    At 30~K, two field-induced transitions are observed, which suggests a change in the zero-field magnetic ordering wave vector [Figure~\ref{MxTxH}(c)]. At 20~K and 1.8~K, a cascade of field-induced transitions is observed with $c$-axis fields [Figures~\ref{MxTxH}(d-e)]. This multistep behavior is a hallmark of a magnetic devil's staircase and strongly suggests a scenario in which distinct exchange interactions compete along the $c$ axis.

\begin{figure*}[!ht]
		\includegraphics[width=0.75\textwidth]{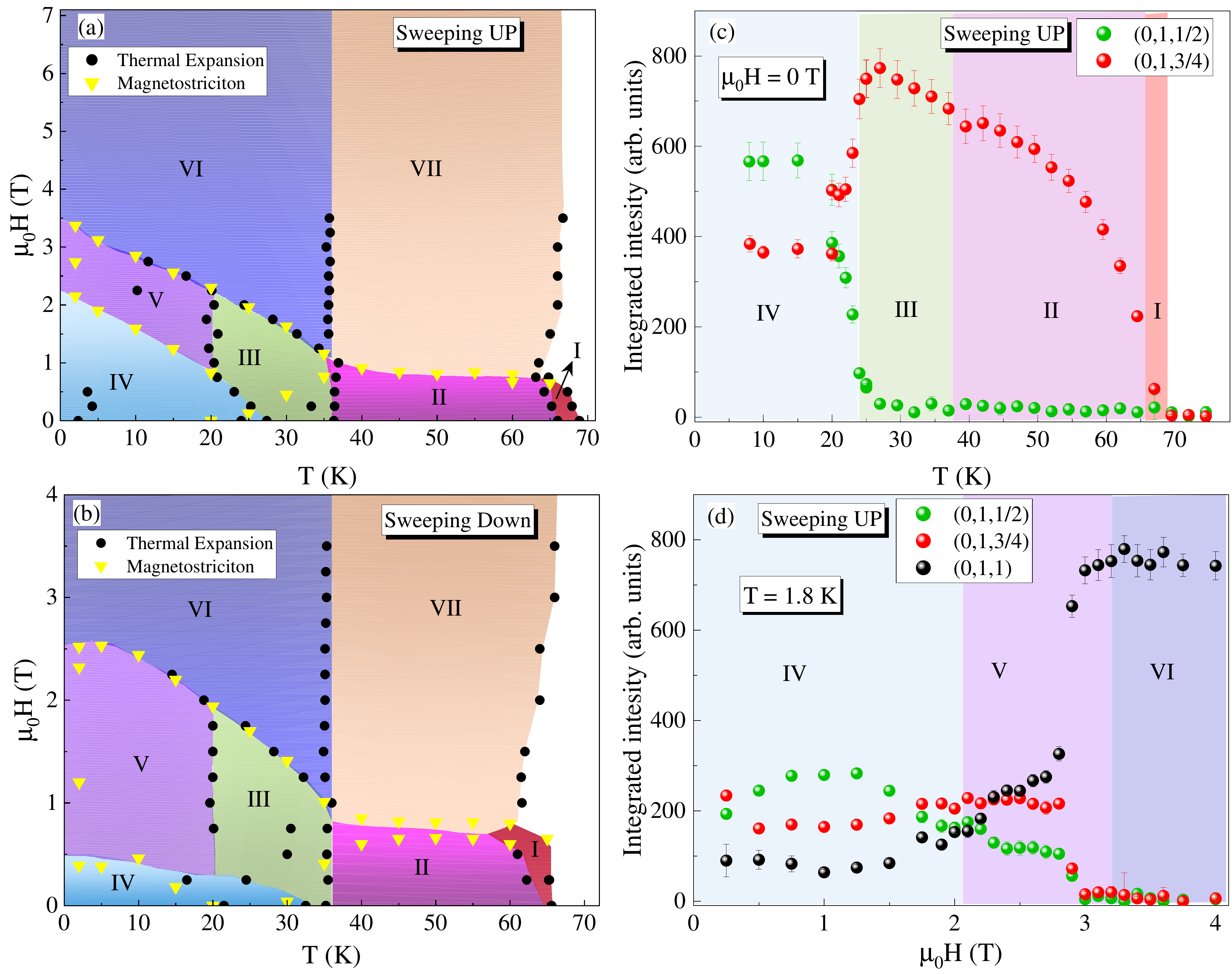}
		\caption{ Proposed phase diagram of UAgBi$_{2}$ constructed with the points extracted from thermal expansion and magnetostriction. (a) the experimental points were extracted from the measurements with temperature and field sweeping up, and (b) were obtained with both sweeping down. (c) Zero-field temperature dependence of the integrated intensity as measured on the magnetic Bragg peaks $(0,1,\frac{1}{2})$ and $(0,1,\frac{3}{4})$. (d) Field dependence at 1.8 K of the magnetic Bragg peaks integrated intensity at $(0,1,\frac{1}{2})$, $(0,1,\frac{3}{4})$, and $(0,1,1)$.  }
		\label{Diagram_neutrons}
	\end{figure*}

    The resulting field-temperature phase diagram for UAgBi$_{2}$ is presented in Figures~\ref{Diagram_neutrons}(a) and (b) for field upsweeps and downsweeps, respectively.
	To build this phase diagram, we also conducted complementary thermodynamic measurements of thermal expansion and magnetostriction (see Supplemental Information for details). At least seven well-defined phases are observed in UAgBi$_{2}$, and magnetic diffraction measurements are essential for a more comprehensive understanding of these phases.


    Single-crystal neutron diffraction at zero magnetic field and 1.8 K reveal the coexistence of two commensurate magnetic propagation vectors $\mathbf{k_1} = (0,0,\frac{1}{2})$ and $\mathbf{k_2} = (0,0,\frac{1}{4})$. Figure~\ref{Diagram_neutrons}(c) shows the temperature dependence of the integrated intensities of the magnetic Bragg peaks at $(0,1,\frac{1}{2})$ and $(0,1,\frac{3}{4})$ upon warming. Clear anomalies at $T_{m5}$, $T_{m4}$, $T_{m3}$, and $T_{m2}$/$T_{m1}$ in the intensity mark the transitions into phases IV, III, and II. These phases are highlighted using the same color scheme as in Figure~\ref{Diagram_neutrons}(a). 

    Phase II, characterized solely by $\mathbf{k_2}$, can be assigned to a commensurate $\downarrow\downarrow\downarrow\uparrow$ magnetic structure, where $\uparrow$ ($\downarrow$) denotes spin-up (spin-down) plane. At phase III below $T_{m3}=36$ K, the integrated intensity at the $(0,1,\frac{3}{4})$ position increases, signaling the enhancement of the $\mathbf{k_2}$ modulation. At lower temperatures, in phase IV, a sharp enhancement of the magnetic Bragg peak associated with the propagation vector $\mathbf{k_1}$ is observed, suggesting a $\uparrow\uparrow\downarrow\downarrow$ type magnetic structure. Notably, the $\downarrow\downarrow\downarrow\uparrow$ and $\uparrow\uparrow\downarrow\downarrow$ magnetic reflections coexist over a finite temperature range, identified as phase IV, indicating a competition between distinct commensurate magnetic states. It is noteworthy that the $\uparrow\uparrow\downarrow\downarrow$ magnetic structure is observed as the ground state in UBi$_2$ and in the antiferromagnetic rare-earth bismuthides CeCuBi$_2$ and CeAuBi$_2$, where ferromagnetic planes are stacked antiferromagnetically along the $c$ axis \cite{adriano_physical_2014,piva_electronic_2020,leciejewicz_neutron-diffraction_1967}.

    We further investigated the magnetic-field dependence of the magnetic Bragg-peak intensities at 1.8 K, as shown in Figure \ref{Diagram_neutrons}(d). The measurements were performed by sweeping the magnetic field up to 4 T after zero-field cooling, allowing direct comparison with the phase diagram in Figure \ref{Diagram_neutrons}(a). Within phase IV, the $\downarrow\downarrow\downarrow\uparrow$ and $\uparrow\uparrow\downarrow\downarrow$ magnetic reflections coexist, consistent with the temperature-dependent measurements, while the applied field induces significant additional k=0 ferromagnetic component, revealed by intensity at $(0,1,1)$.


    At the boundary between phases IV and V, the field-dependent integrated intensities of the magnetic Bragg peaks overlap and cross, providing clear evidence of strong competition between distinct magnetic states. Importantly, this behavior is directly correlated with the appearance of multiple steps in the magnetization data, demonstrating that the stepped magnetization is not incidental but reflects an intrinsically complex magnetic energy landscape in this field range. Within phase V, the intensity associated with the $\downarrow\downarrow\uparrow\uparrow$ structure is progressively suppressed, while the $\uparrow\uparrow\uparrow\downarrow$ phase 
    becomes dominant. At 1.8 K and above 3 T, the Bragg reflections corresponding to the $k=(0,0,\frac{1}{2})$ and $k=(0,0,\frac{1}{4})$ phases vanish, while intensity of measured k = (0, 0, 0) reflections increases. We note, however, that we cannot rule out the presence of additional reflections within phase VI that are beyond our current reciprocal space scans. Therefore, we cannot ascertain whether phase VI is a fully polarized phase.





\begin{figure}[!ht]
	\includegraphics[width=1\columnwidth]{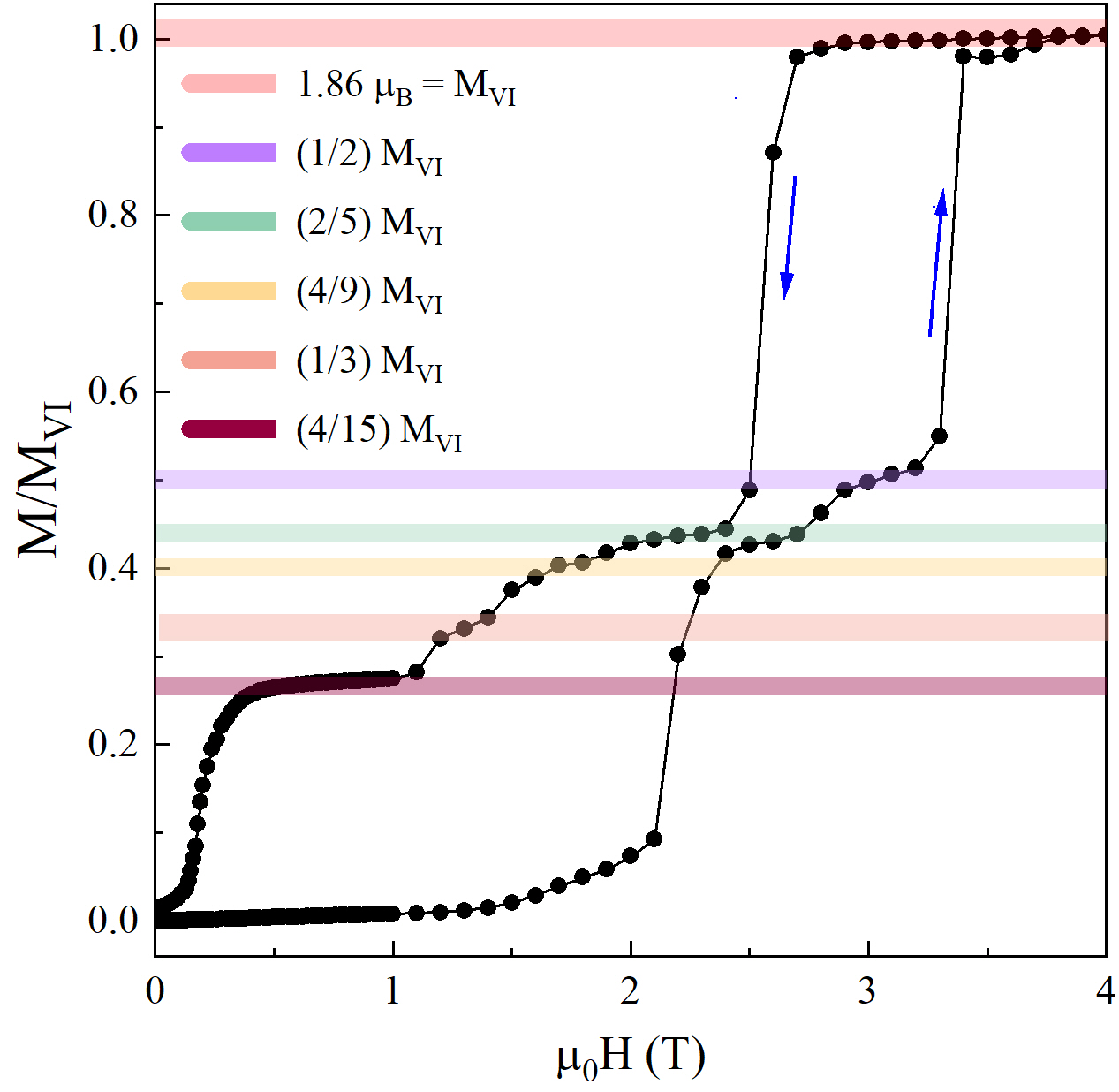}
	\caption{  Zoom-in of the magnetization at 1.8 K as a function of magnetic field applied along the $c$ axis, showing the magnetization steps expressed as the fractional magnetization $M/M_{VI}$.   }
	\label{Devil}
    \end{figure}

    Assuming that phase VI exhibits a maximum magnetization of $M_{VI}=1.86~\mu_B$ [Figure~\ref{MxTxH}(e)], we examine the fractional values of the magnetization plateaus by plotting $M/M_{VI}$ as a function of magnetic field applied along the $c$ axis. Figure~\ref{Devil} reveals at least five distinct magnetization steps at fractional values of $\frac{1}{2}M_{VI}$, $\frac{2}{5}M_{VI}$, $\frac{4}{9}M_{VI}$, $\frac{1}{3}M_{VI}$, and $\frac{4}{15}M_{VI}$. Notably, the appearance of these steps depends strongly on the magnetic-field history. For instance, steps at $\frac{2}{5}M_{VI}$ and $\frac{1}{2}M_{VI}$ are observed during field upsweeps, whereas steps at $\frac{2}{5}M_{VI}$, $\frac{4}{9}M_{VI}$, $\frac{1}{3}M_{VI}$, and $\frac{4}{15}M_{VI}$ emerge only during downsweeps. Assuming that these fractional magnetization states correspond to spin configurations with moments constrained along the $c$ axis, as supported by magnetization and neutron diffraction measurements, the field-induced spin arrangements proposed in Figure~\ref{Devil} provide a consistent description of the experimental observations.

    The presence of multiple zero-field phases in UAgBi$_{2}$, combined with the field-induced step-like features in the magnetization, closely resembles the behavior seen in MDS compounds. We now turn to the outstanding question of why MDS behavior is not widely observed in uranium-based materials. First, MDS behavior relies on the delicate balance between an easy-axis anisotropy that is strong enough to align moments along one, unique spatial axis but not too strong such that the degeneracy of magnetic configurations is completely lifted. In the case of CeSb, this degeneracy is closely related to the mixing between the 4\textit{f} CEF $\Gamma_7$ ground state and its 4\textit{f}$\Gamma_8$ excited state, supported by the small energy splitting of 37~K between them \cite{kuroda_devils_2020, xu_orbital-flop_2019, heer_neutron_1979, kioussis_mechanism_1988}. In fact, the overall CEF splitting in UAgBi$_{2}$ is the smallest of all U$M$Bi$_{2}$ members~\cite{freitas2025_arxiv}, which in turn generates the smallest magnetic anisotropy. 

	Further insights into the importance of CEF effects come from a comparison to UAuBi$_2$. For instance, UAgBi$_2$ has a reduced magnetic anisotropy compared to UAuBi$_2$: the anisotropy ratio $\chi_{||}/\chi_{\perp}$ is only 50 at $T_{m5}$ in UAgBi$_{2}$, whereas it reaches 135 at $T_C = 22$~K in UAuBi$_{2}$. Enhanced magnetic anisotropy typically indicates a larger overall CEF splitting, and the overall CEF splitting in UAuBi$_{2}$ is indeed large ($\sim 4800$~K), as determined from fits to the magnetic susceptibility \cite{rosa_ferromagnetic_2015}. To estimate the overall splitting in UAgBi$_{2}$, we calculate the leading CEF parameter \cite{hutchings_point-charge_1964}, $B_2^{0}$, using the high-temperature expansion of the magnetic susceptibility \cite{ wang_crystal-field_1971}:
	\begin{center}
		\begin{equation}
			B_2^{0} = \frac{10(\theta{\perp} - \theta_{||})}{3(2J-1)(2J+3)}.
			\label{eq3}
		\end{equation}
	\end{center}

	$B_2^{0}$ is proportional to the difference in Weiss temperatures along different axes. In general, the larger the magnetic anisotropy, the higher the magnitude of the $B_2^{0}$ parameter, and consequently, the larger the overall splitting of the CEF levels. For UAuBi$_{2}$, the derived $B_2^{0}$ is -4.27 K \cite{rosa_ferromagnetic_2015}, whereas $B_2^{0}=-2.5$~K for UAgBi$_{2}$, which again suggests a smaller overall CEF splitting in UAgBi$_{2}$. A more comprehensive investigation of CEF effects in UAgBi$_{2}$, including fits to the magnetic susceptibility, confirms the reduced overall CEF splitting and the presence of low-lying CEF states~\cite{freitas2025_arxiv}. Low-lying CEF levels typically increase the number of nearly-degenerate ground states and are generally expected to enhance magnetic frustration. This expectation arises because excited CEF levels typically introduce more spin and orbital degrees of freedom, $i.e.$, there may not be a single dominant direction along which the magnetic moments prefer to be aligned. 


    Moreover, the exchange interactions along the $c$ axis must be finely tuned such that two dominant interactions of opposite sign are present. Although this balance depends on many parameters and is difficult to predict, one can reasonably expect that, in the simplest cases, the exchange interactions are largely governed by the overlap of electronic orbitals, either directly or indirectly via the Fermi surface, between the magnetic ions in the crystal structure. This overlap is intimately related to the spacing between uranium moments along the $c$ axis, which in turn depends on the transition-metal layer. In the 112 series of compounds, the additional \textit{M} layer intercalated between the $R$Bi$_2$ ($R=$ Ce, U) unit cells provides exceptional chemical tunability \cite{kaczorowski_structural_1992,rosa_ferromagnetic_2015}. Within Ce members, the introduction of Cu and Au layers enhances AFM magnetic order from $T_{N}=3.3$~K in CeBi$_{2}$ to $T_{N}=16$~K in CeCuBi$_2$ and $T_{N}=19$~K in CeAuBi$_2$, and the $M=$ Au member shows larger magnetic anisotropy compared its Cu counterpart. A simple A-type magnetic structure ($\uparrow\uparrow\downarrow\downarrow$) is realized in these members, and magnetic moments are confined along the $c$ axis \cite{adriano_physical_2014, piva_electronic_2020}. In contrast, CeAgBi$_2$ displays the smallest transition temperature in this family ($T_N=6.4$~K), a significantly reduced magnetic anisotropy compared to its Cu- and Au-based analogs, and a complex field-temperature phase diagram with a cascade of field-induced transitions \cite{thomas_hall_2016}. 

    UBi$_{2}$, which crystallizes in the same $P4/nmm$ space group, orders antiferromagnetically below $180.8$~K with moments along the \textit{c} axis in a $\uparrow\downarrow$ configuration \cite{leciejewicz_neutron-diffraction_1967}. 
    In contrast to its Ce counterparts, the introduction of Cu(3\textit{d}) layers significantly reduces $T_N$ to 51~K in UCuBi$_2$, with a subsequent AFM transition at 15~K \cite{kaczorowski_structural_1992}. Conversely, Au(5\textit{d}) layers in UAuBi$_{2}$ promote ferromagnetism below $T_{c} =22.5$~K, accompanied by substantial $c$-axis anisotropy \cite{rosa_ferromagnetic_2015}. We therefore conclude that, in addition to sharing the same crystalline structure, Ce\textit{M}Bi$_2$ and U\textit{M}Bi$_2$ compounds also share the non-monotonic evolution of magnetic properties and anisotropy with isovalent chemical substitution from a 3\textit{d-} to a 4\textit{d-} to a 5\textit{d-} transition metal. We note that such non-monotonic evolution is also observed more broadly in other families of ternary intermetallic compounds with distinct ground state properties, such as the Ce(Co, Rh, Ir)In$_5$ series (Ce115). In this series, the Co(3\textit{d}) and Ir(5\textit{d}) members exhibit a heavy-fermion superconducting ground state with $T_c$ values of 2.3~K and 0.4~K, respectively \cite{petrovic_heavy-fermion_2001,petrovic_new_2001}. In contrast, the Rh(4\textit{d}) member shows a spiral AFM state below $T_{N} = 3.4$~K \cite{bao_incommensurate_2000,hegger_pressure-induced_2000}. The 4\textit{d} compounds in both 112 and 115 series exhibit the most complex properties compared to their analogs, which suggests that the non-monotonic chemistry details of the 4\textit{d} orbitals (e.g., atomic radius, position of $4d$ bands with respect to the Fermi level, shielding effects) play an important role, as proposed previously by DFT+DMFT calculations in the Ce115 materials \cite{haule_dynamical_2010}.

	In summary, our experimental investigation of newly-discovered compound UAgBi$_{2}$ reveals at least four distinct zero-field magnetically ordered states as a function of temperature and a cascade of field-induced transitions when fields are applied along the easy $c$ axis, a hallmark of magnetic devil's staircase behavior. Our magnetic susceptibility results point to the presence of low-lying crystal field levels, which contribute to magnetic frustration and nearly degenerate states. Our findings not only highlight the importance of the transition metal layer on magnetic frustration, but they also provide a platform for further investigation of devil's staircase phenomena in $5f$-based materials.

	
	\section{Acknowledgments}

	Work at Los Alamos was supported by the U.S. Department of Energy, Office of Basic Energy Sciences, Division of Materials Science and Engineering project ``Quantum Fluctuations in Narrow Band Systems''. C. G. acknowledge support from the Laboratory Directed Research and Development (LDRD) program. C.S.T.K. gratefully acknowledges the support of the U.S. Department of Energy through the LANL/LDRD Program and the G. T. Seaborg Institute.
    Scanning electron microscope and energy dispersive x-ray measurements were performed at the Electron Microscopy Lab and supported by the Center for Integrated Nanotechnologies, an Office of Science User Facility operated for the U.S. Department of Energy Office of Science. GSF acknowledges the National High Magnetic Field Laboratory, where a portion of this work was performed. The National High Magnetic Field Laboratory is supported by the National Science Foundation through Cooperative Agreement No. DMR-2128556, the State of Florida, and the U.S. Department of Energy.
    GSF and PGP acknowledge the support from FAPESP (Grants No.  2017/10581-1, 2019/26247-9, 2022/09240-3) and CNPq (Grants No. 140724/2019-2, 405408/2023-4, and 311783/2021-0. We further acknowledge support from the INCT project Advanced Quantum Materials, involving the Brazilian agencies CNPq (Proc. 408766/2024-7), FAPESP (Proc. 2025/27091-3) and CAPES.
    DRY and MJ acknowledge funding by the Swiss National Science Foundation through the project “Berry-Phase Tuning in Heavy f-Electron Metals (\#200650)”. 
    DRY and WS were supported through funding from the European Union’s Horizon 2020 research and innovation programme under the Marie Sklodowska-Curie grant agreement No 884104 (PSI-FELLOW-III-3i). 
    This work is based on experiments performed at the Swiss ZEBRA single crystal diffractometer spallation neutron source SINQ, Paul Scherrer Institute, Villigen, Switzerland. 
    KK acknowledges funding by the Swiss National Science Foundation (Grant No. 200021-219950).
	
    \appendix
    \section{Experimental details}
	
	Single crystals of UAgBi$_{2}$ were grown by a combined (Ag+Bi) self-flux technique with an initial composition of U:Ag:Bi = 1:10:10. The reagents were placed in an alumina crucible and vacuum sealed in a quartz tube, which was subsequently heated to 1050 $^{\circ}$C for 8 h. The solution was then cooled at a rate of 5$^{\circ}$C/h to 550 $^{\circ}$C, and the Bi excess was removed by centrifugation. Plate-like crystals were mechanically removed from the crucible.

		Magnetic susceptibility and magnetization measurements were carried out using a Quantum Design Magnetic Properties System (MPMS3) equipped with a 7 T magnet. Temperature-dependent magnetic susceptibility measurements were performed with different applied magnetic fields parallel and perpendicular to the \textit{c} axis. Magnetization measurements at constant temperatures were  conducted from 0 to 7~T for both crystallographic directions. Specific heat and resistivity measurements were performed in a Quantum Design Physical Property Measurement System (PPMS) equipped with 9 T and 16 T magnets and a $^{3}$He system that can reach 0.4~K. 
		
		Neutron diffraction measurements were performed on a $0.18$~g single crystal of UAgBi$_2$ using the thermal single crystal diffractometer ZEBRA at SINQ (PSI). The sample was aligned with the $c$ axis vertical, and cooled in a vertical 6 T magnet. Measurements were performed in the normal beam geometry using a neutron wavelength $2.3$~\AA, selected with a pyrolytic graphite (002) monochromator, and detected using either a $^3$He tube detector or area detectors.

        \section{Crystal structure refinement}
        Room temperature single crystal X-ray diffraction measurements were performed on a Bruker D8 Venture diffractometer with Mo K$\alpha$ radiation. The tetragonal crystalline structure of the P4/nmm space group was determined using APEX 3 software. The obtained lattice and atomic parameters are shown in Table \ref{table2}.

        	\begin{table}[!h]
		\caption{Structural refinement parameters at room temperature collected on single crystals of UAgBi$_2$.}
		\label{table2}
		
		\begin{tabular}{ccccccccccccccccccccccccccccccccccccccccccc} \hline \hline
			\noalign{\vskip 1mm}
			&& && && && && && &&  a = b   && && && && && && && 4.5067(6)\AA && && && && && && &&\\
			&& && && && && && && c   && && && && && && && 10.3027(13)\AA && && && && && && &&\\
			
		\end{tabular}
		
		\begin{tabular}{ccccccccccccc} 
			\noalign{\vskip 1mm}
			Atom && Wyckoff && x && y && z && $U_{iso}$ && Occupancy \\
			\tableline
			\noalign{\vskip 1mm} U && 2c && 1/4 && 1/4 && 0.24132(4) && 0.0099(1) && 1 \\
			Ag && 2b && 3/4 && 1/4 && 1/2 && 0.0160(3) && 1 \\
			Bi && 2c && 1/4 && 1/4 && 0.68649(4) && 0.0112(1) && 1 \\
			Bi && 2a && 3/4 && 1/4 && 0 && 0.0102(1) && 1 \\	
			\hline \hline
		\end{tabular}
	\end{table}

	\section{Thermal expansion and magnetostriction}
\begin{figure}[h!]
	\begin{center}
		\includegraphics[width=1\columnwidth]{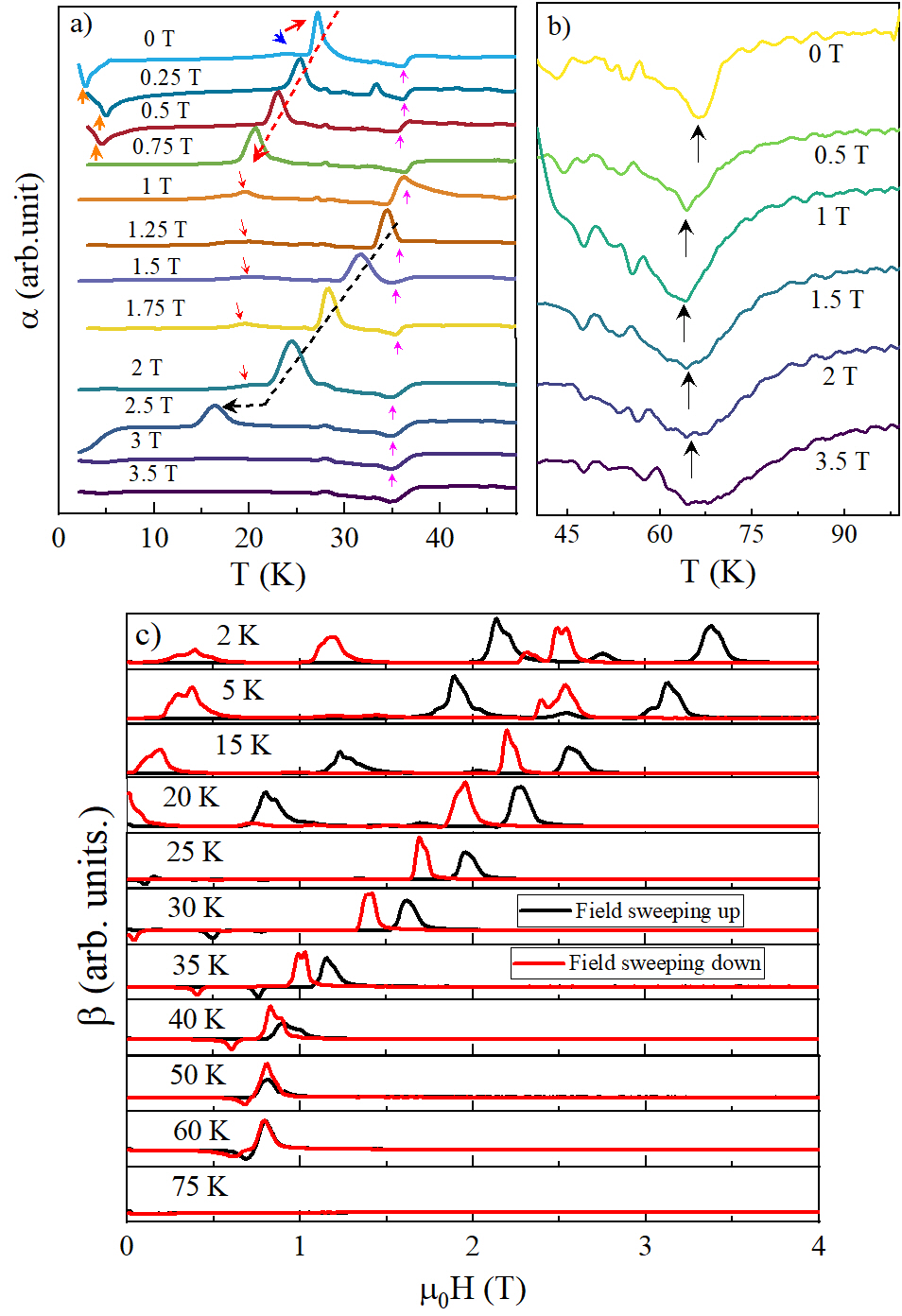}
		
	\end{center}	
	\caption{  a) and b) Magnetic field dependence of the linear thermal expansion coefficient parallel to the c axis tracking the evolution of the magnetic transitions. We have shifted the curves and drawn arrows to facilitate the tracking of each transition. c) Temperature dependence of linear magnetostriction coefficient measured with an external magnetic field and length change parallel to the c axis. }
	\label{Ther_Mag}
\end{figure}

    The thermal expansion and magnetostriction measurements performed are shown in Figure \ref{Ther_Mag}. We present the temperature dependence of the linear thermal expansion coefficient ($\alpha$) and magnetostriction coefficient ($\beta$) at different applied magnetic fields parallel to the c axis. The length variation ($\Delta L$) was measured along the c axis. At zero field, one can observe the same anomalies as observed in the specific heat measurements by the drawn arrows. Furthermore, as the applied magnetic field increases, these anomalies behave similarly to those observed in $c_{p}(T)$ and $M(T)$. The field dependence of the magnetostriction coefficient ($\beta$) displays field-induced transitions similar to those observed in the $M(H)$ data. 

\bibliography{Ref_UAgBi2}

\end{document}